\documentclass[11pt,twoside]{article}
\usepackage{asp2010}

\resetcounters

\begin{document}

\title{Long Period Variables in the Magellanic Clouds with EROS-2 survey}
\author{Maxime~Spano,$^1$ Nami~Mowlavi,$^{1,2}$ Laurent~Eyer,$^1$
Jean-Baptiste~Marquette,$^{3,4}$\\ and Gilbert~Burki$^1$
\affil{$^1$Observatoire de Gen\`eve, Universit\'e de Gen\`eve, 51 Chemin des Maillettes, 1290 Sauverny, Switzerland\\
$^2$ISDC, Observatoire de Gen\`eve, Universit\'e de Gen\`eve, 1290 Versoix, Switzerland\\
$^3$UPMC, Universit\'e Paris 06, UMR7095, Institut d'Astrophysique de Paris, F-75014, Paris, France\\
$^4$CNRS, UMR7095, Institut d'Astrophysique de Paris, F-75014, Paris, France\\}}

\begin{abstract}
We use the EROS-2 database to analyze Long Period Variables (LPVs) in the Magellanic Clouds. It results in the creation of a catalog of more than 40\,000 LPV candidates which is briefly introduced here.
\end{abstract}

The EROS-2 survey aimed at searching microlensing events in two non standard $B_{EROS}$ (420 -- 720 nm) and $R_{EROS}$ (620 -- 920 nm) bands from July 1996 to March 2003. The targets were the Large Magellanic Cloud (LMC), the Small Magellanic Cloud (SMC), the Galactic center, and the Galactic spiral arms. It resulted in a database of more than 87 million sources suitable for the study of variable stars.

The selection of LPV candidates is based on the smoothness of the light curves, evaluated by applying the statistical Abbe test \citep{vonNeumann41}, and on the color of the star, requiring the star to be redder than the red clump.
Our resulting sample contains more than 38\,000 LPVs for the LMC, and more than 4\,000 LPVs for the SMC. Multiple periods are searched for with an iterative pre-whitening procedure that successively extracts significant periods in the Deeming periodogram, the peak being considered significant if higher than at least 3 times the standard deviation of the periodogram. Cross-matching with the 2MASS catalogue, period-magnitude relations are constructed using the 2MASS $K_{S}$ magnitudes ($\sim$97\% crossmatches found for both Clouds).

\begin{figure}[!ht]
\plottwo{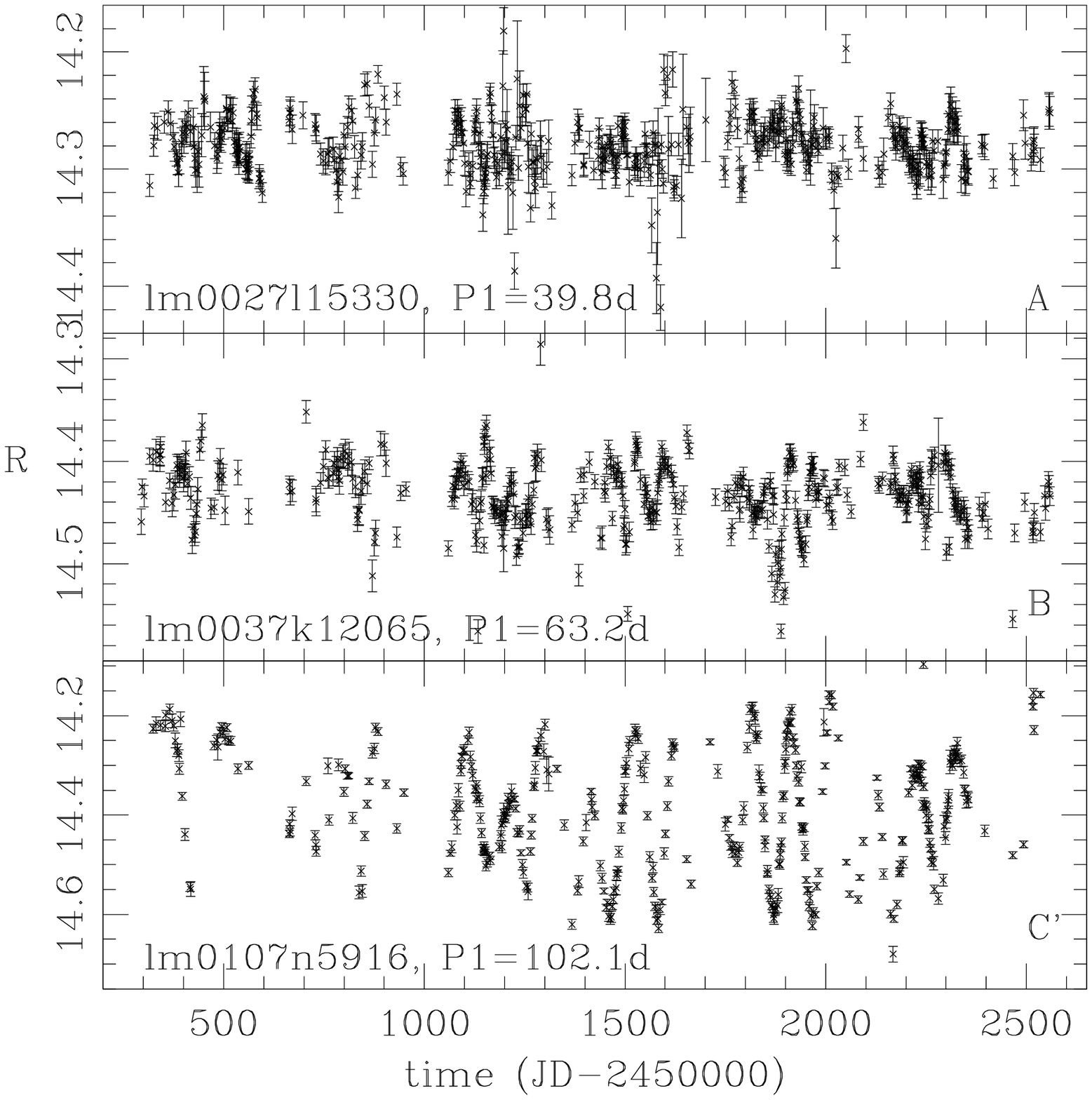}{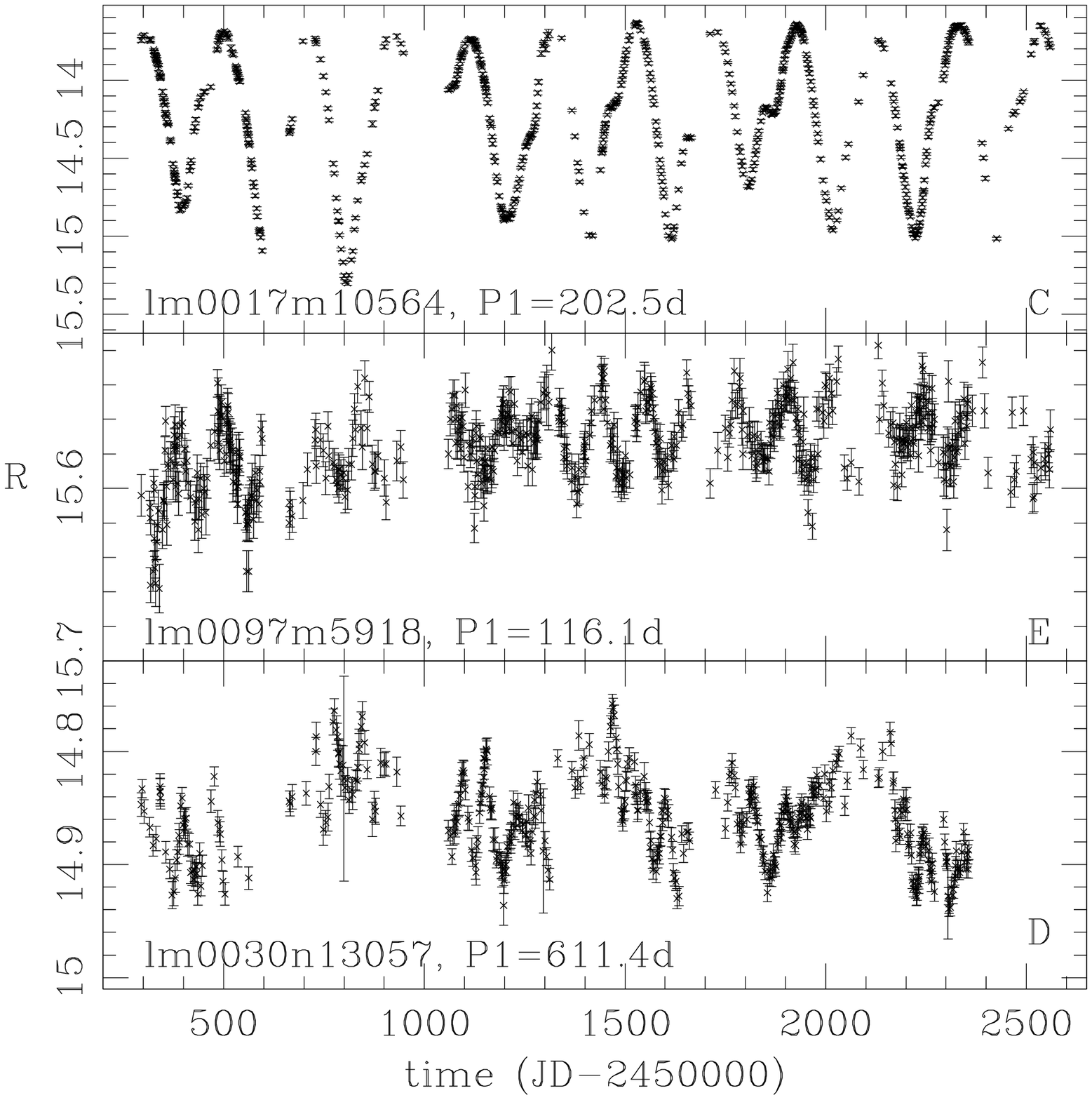}
\caption{Examples of light curves in the R band of LPV candidates from the LMC lying in sequences A to D. The corresponding most significant periods are indicated, in days.}
\label{figure1}
\end{figure}

The traditional period-magnitude sequences, usually named A, B, C', C, E and D from the shortest to the longest periods (cf. \citet{Woodetal99}, \citet{Wood00}, \citet{Itaetal04}), are found. Examples of typical light curves for each sequence are shown in Fig.~\ref{figure1}. The light curves become more regular from sequence A to C, the latter sequence harboring the fundamental mode pulsators. Variables from sequence E are populated by binaries, and sequence D contains the long secondary period variables, the physical process of which is still unknown (\citet{Nichollsetal10}, see also Nicholls, in these proceedings). The sequences appear similar for the SMC and LMC among our LPV candidates.

Period-color relations are constructed using 2MASS J and  $K_{S}$ magnitudes. We find a change of the slope of the relations according to the J- $K_{S}$ color, variables in the range $1.4<J-K_{S}<2$, populated by C-rich AGB stars, producing sequences with a steeper slope than the ones with a $J-K_{S}<1.4$ composed by O-rich AGB stars.

The catalogue will be presented and made available in Spano et al. (2010a and 2010b, in prep.), together with an analysis of the results. It will contain the main properties of the LPVs such as periods, amplitudes, colors, and magnitudes.
A study of the EROS-2 data from the Galactic center and spiral arms will also be presented in Spano et al. (2010b, in prep.), extending the present study to higher metallicities.

\acknowledgements This publication makes use of data products from the Two Micron All Sky Survey, which is a joint project of the University of Massachusetts and the Infrared Processing and Analysis Center/California Institute of Technology, funded by the National Aeronautics and Space Administration and the National Science Foundation.\\
It also makes use of EROS-2 data which were kindly provided by the EROS collaboration. The EROS (Exp\'erience de Recherche d\textquoteright Objets Sombres) project was funded by the CEA, the IN2P3 and INSU CNRS institutes.

\bibliography{Spano_author}

\end{document}